\documentclass[paper]{aa}
\usepackage{graphicx}
\usepackage[english]{babel}
\usepackage{amsmath}
\usepackage{amssymb}
\usepackage{txfonts}
\begin{document}
   \title{The stellar population of the Rosat North Ecliptic Pole survey\\ II. Spectral analysis}

   \titlerunning{The stellar North Ecliptic Pole Survey}

   \author{L. Affer
          \inst{1,2}
          \and
          G. Micela
          \inst{1}
	  \and
          T. Morel
          \inst{3}     }

   \offprints{affer@astropa.inaf.it}
 
 \institute{Istituto Nazionale di Astrofisica, Osservatorio Astronomico di Palermo G.\,S. Vaiana, Piazza del Parlamento 1, 90134 Palermo, Italy
	  \and
	  Dipartimento di Scienze Fisiche ed Astronomiche - Universit\`a di Palermo, Piazza del Parlamento 1, 90134 Palermo, Italy
	  \and   
	     Katholieke Universiteit Leuven, Departement Natuurkunde en Sterrenkunde, Instituut
	     voor Sterrenkunde, Celestijnenlaan 200D, 3001 Leuven, Belgium}

    \date{Received; Accepted}

    \abstract
    {X-ray surveys allow to identify young, main-sequence stars in the solar neighborhood. Young,
    stellar samples, selected according to their activity, can be used to determine the stellar birthrate in the last billion years. 
    The \textit{ROSAT} North Ecliptic Pole survey (NEP), with its moderately deep sensitivity (fluxes $\approx 10^{-14}$ erg cm$^{-2}$\, sec$^{-1}$), is the best survey, to date, able to sample the
    intermediate-age (10$^8$ - 10$^9$ years) nearby population. 
    The identification process of NEP
    X-ray sources resulted in 144 X-ray sources having a normal stellar counterpart, with an
    excess of yellow stars with respect to model predictions.}
    {We want to determine if these X-ray active stars are young or intermediate-age
    stars, or active binaries.}
    {We acquired high-resolution, high signal-to-noise ratio optical spectra, to determine the age and physical properties of the NEP X-ray-detected stellar sources. We measure the (i) lithium abundance using the \ion{Li}{i} 6707.8 \AA\, line, which is an excellent, youth indicator for our age range of interest; (ii) rotational and radial velocities (through cross-correlation methods); and (iii) chromospheric emission (from H$\alpha$ and {\ion{Na}{i}} $D{_1}$ and $D{_2}$ lines).}
    {The radial velocities distribution is consistent with that of a young field star population
    of age 4x10${^9}$ yrs, or younger. Rotational velocity measurements
    imply that our sample is dominated by
    relatively young or intermediate-age stars, as confirmed by our lithium measurements.}
    {Most of the detected stars 
    probably belong to a young or intermediate-age population. 
    Our measurements suggest that a burst in the stellar birthrate of a factor of four occurred in the last 10$^8$ years. We cannot, however, exclude the possibility that a small fraction of sources, amongst the fastest of the K-rotators, are old binary systems with tidally-locked rotation.}

    \keywords{stars: rotation - stars: abundances - techniques: spectroscopic - techniques: radial velocities - Galaxy: solar neighbourhood}

   \maketitle
%

\section{Introduction} \label{sect_intro}
The analysis of the stellar component of flux-limited X-ray surveys is a
powerful tool in the study of the young stellar population of the Galaxy. 
Optical data, which helps determine photospheric parameters, cannot alone discriminate between main-sequence stars of different age.
In contrast, the average X-ray luminosity of normal, late-type stars decreases by up to 3 orders of magnitude mainly 
during the main-sequence lifetime (Micela et al. 1985, 1988; Guedel et al. 1997;
Feigelson \& Montmerle 1999); optical properties, however, remain almost unchanged. 
Stellar X-ray surveys allow to study in detail the global properties of stellar populations in the solar
neighborhood, in particular of the young population, since young stars are detected, in X-rays,
out to larger distances than older stars. 
Shallow stellar, flux-limited, X-ray selected samples will be dominated by young stars; in contrast, deep high-latitude stellar X-ray samples will be dominated by old stars, because young stars have smaller scale height, and deep X-ray surveys probe distances beyond the scale height of the youngest population (Micela et al. 2001).
The comparison between observations and predictions from galactic models offers an important tool to constrain the spatial distribution of young stellar population and to infer the local stellar birthrate in the last billion years. 

Analyses of both the \textit{Einstein} Extended Medium Sensitivity
Survey (EMSS, Gioia et al. 1990), and  \textit{XMM-Newton} Bright Serendipitous Survey (XBSS, Della Ceca et al.
2004) measured an excess of yellow stars (Sciortino et al.
1995; Lopez-Santiago et al. 2006, respectively). For the EMSS, these yellow
stars appear to be young, in agreement with their measurement of lithium abundance (Favata et al. 1993). For the XBSS,
the excess may represent a young population, or a number of active binary
systems that have a yellow dwarf as a primary (Lopez-Santiago et al. 2007). 

The study of the stellar content of the
\textit{ROSAT} North Ecliptic Pole survey (NEP hereafter, Henry et al. 2001) confirmed the existence of an excess of yellow stars in the solar neighborhood even at intermediate X-ray flux, which is probably due to a young stellar population (Micela et al. 2007, hereafter referred to as \ion{Paper}{i}).

The NEP survey is a survey of X-ray sources, covering a 9${^\circ}\times 9{^\circ}$ area centred on the Galactic
North Ecliptic Pole ($l=96{^\circ}.4$, $b=29{^\circ}.8$). This is the sky region with the highest sensitivity observed
during the \textit{ROSAT} All Sky Survey (RASS), with a total exposure time amounting to $\sim$\, 40 ks at the pole.
The relatively high galactic latitude, together with the moderate sensitivity, permits to observe young stars close to their scale height.

An overview of the NEP survey, including both its selection function and optical identification
program, can be found in Henry et al. (2001); Voges et al. (2001) summarized the X-ray data and
source statistical properties; Gioia et al. (2001) provided evidence for cluster X-ray luminosity
evolution, and Mullis et al. (2001) described the NEP supercluster discovered in the survey X-ray data. The
extragalactic component of the NEP survey was discussed in Mullis et al. (2003, 2004a,b), Gioia et al.
(2004), and Henry et al. (2006).

The identification of the optical counterparts (based both on optical photometry and
spectroscopy of all candidate counterparts) to each of the X-ray sources was completed by Gioia et al. (2003). This X-ray based data set is
unique in its combination of contiguous sky coverage (80 deg$^2$), and sensitivity, which is higher than any of the \textit{Einstein}-based surveys.

More recent X-ray surveys, such as the \textit{Chandra} Deep Fields, are far deeper than the NEP survey, but have small sky coverage. These surveys are biased towards older stars, which are more numerous but fainter in X-rays, with the younger stellar population being very poorly sampled. The small sky coverage does not, however, allow robust conclusions to be made.

The stellar content of the \textit{Chandra} deep survey (Feigelson et
al. 2004) reveals that the old star population does not contain any excess of yellow stars, but
instead a deficiency of dG-dK stars, while the signatures of dM stars are reproduced perfectly by galactic models.
This behaviour is the opposite of that observed for the young stellar population, and suggests that some changes occur at intermediate ages, where NEP survey may give a unique contribution thanks to its intermediate
sensitivity. Thus the NEP survey is unique in its ability to properly sample the young and intermediate-age stellar populations in the Galactic disk. 

In \ion{Paper}{i} we analyzed the stellar content of the NEP survey, to determine the nature of active stars in the solar neighborhood. 
We determined the spectral types for all source detections. We then compared the observational data with the predictions of the XCOUNT model (Favata et al. 1992; Micela et al. 1993), and, at intermediate X-ray fluxes, ascertained the presence of an excess of active yellow stars, in the solar neighborhood. 

To characterize the stellar population of the NEP survey and determine the nature of the observed excess, we derived, in the second paper of this series, the physical properties of the X-ray sources, such as radial and rotational velocities, and lithium abundances.
In a forthcoming paper, we will determine the chemical composition of the stellar
photospheres; this will allow the study of the enrichment history of the interstellar matter and,
combined with kinematical data and ages, will provide a powerful way of probing the chemical and
dynamical evolution of the Galaxy.
 
The present paper is organized as follows: in Sects. 2 and 3, we describe the selection rules of our sample,
the observations, and the data reduction. In Sect. 4, we describe the methods of analysis and discuss the distribution of radial and rotational velocities, while binary stars and lithium-abundance determination are described in Sect. 5 and 6. In Sects. 7 and 8, we discuss and summarize our findings.

\begin{table}
\caption{Sampled stars per spectral type}
\begin{center}
\label{tab0}
\begin{tabular}{rcc} \hline
Sp.Type	&	N.Obs.(\ion{Paper}{i})&	N.Obs.(This work)\\\hline
A&	3&	1\\
F0-F5&	10&	9\\
F6-F9&	8&	4\\
G&	29&	18\\
K&	53&	22\\
M&	41&	2\\\hline
Total	&	144&	56\\\hline
\end{tabular}
\begin{flushleft}
\end{flushleft}
\end{center}
\end{table} 

\section{Observations and data reduction} \label{sect_obs}

To complete our proposed investigation, we would have ideally wanted to observe the entire sample of NEP stars. 
Due to telescope and instrumentation constraints we instead had to
restrict our sample to stars with $J\le11$ in the 2MASS Catalogue; 
this selection criterion should not affect the achievement of our goals, apart from the study of dM stars. 
In Table~\ref{tab0}, we summarize the spectral type classification of stellar X-ray NEP sources, as derived in \ion{Paper}{i}, together with the sample observed in the present work. As can be seen from Table~\ref{tab0}, we covered almost the entire sample of early-F (F0-F5) stars, while we sampled a fraction between 40\% and 60\% of late-F to G
and K stars; dM stars were poorly sampled (due to their faintness at the energies considered) and we are unable to measure reliably their mean properties, even though, with dK stars, they form a substantial part of the data set.
Finally we can state that our sample is representative of stars earlier than dM. 

The observations, with signal-to-noise ratios between 40 and 440, were obtained (in service mode) during several
observing runs between January and June 2004 (see Table~\ref{tabA}) at the Telescopio Nazionale
Galileo (TNG) located at the Observatorio del Roque de Los Muchachos (La
Palma, Canary Islands). 

Our survey contains 56 stars in the region of sky of 80.6 deg${^2}$ about the North Ecliptic Pole (NEP) at
$\alpha$=$18{^h}00{^m}00{^s}$, $\delta$=$+66{^\circ}33'00''$ (J2000.0), at a
Galactic latitude, $b=29.{^\circ}8$, with spectral types in the range F0V-M3V, and values of $B-V$ in the range
0.3 to 1.5, which corresponds approximately to effective temperatures of about 7300 K-3500 K.
For all observations, we used
the cross-dispersed echelle SARG spectrograph attached to the 3.58 m TNG. The dioptric camera has a focal ratio of $\it{f}$/4.8, and the detector is a mosaic of 2 2k $\times$ 4k
thinned and back illuminated CCDs, with a 13.5 $\mu$m pixel size, and a scale of 0.16
arcsec/pixel. We used the Yellow grism, with a dispersion of 61.0 \AA/mm, the FW4 filter, and selected a binning of 2$\times$1. The spectra continuously cover a wavelength
range from 4620 \AA\, to 7920 \AA, with a resolving power of about 57000, yielding a FWHM of 5 km s$^{-1}$. 

To enable a differential analysis, we obtained a solar flux spectrum,
i.e. a high S/N ($\approx$ 500) moonlight spectrum acquired using the same instrumental configuration
as the target stars.

As preparation for our cross-correlation analysis, the objective of the data reduction is to produce wavelength-calibrated, one-dimensional spectra for
each echelle order, free of relevant instrumental and telluric signatures.
Tasks, implemented in the {\tt IRAF}\footnote{{\tt IRAF} (Image Reduction and Analysis
Facility) is distributed by the National 
Optical Astronomy Observatories, operated by the Association of Universities for Research 
in Astronomy, Inc., under cooperative agreement with the National Science Foundation.} 
package, were used to carry out the standard reduction procedures on echelle spectra 
(i.e. bias subtraction, flat-field correction, removal of scattered light, order extraction, 
and wavelength calibration). The two CCDs were separated and reduced independently. 
Numerous emission spikes, possibly arising from CCD electronics, were initially 
found to affect the extracted spectra. To remedy this problem, the data were extracted 
after excluding deviant pixels in the two-dimensional science frames. 
The rejection thresholds of the sigma-clipping algorithm were varied to remove the artefacts efficiently, without affecting significantly the remainder of the spectrum.
The wavelength calibration was performed using Thorium-Argon lamp exposures acquired each night. 
For three nights, Thorium-Argon lamp exposures were not acquired; for data obtained during this time, wavelength calibration was completed using lamp exposures available for the night closest in time; we identify the spectra affected by labelling the corresponding stars with an asterisk in Table~\ref{tabA} (we use a similar symbol in all tables). We compared the wavelength shifts between telluric lines of these
spectra, and of spectra for which wavelength calibration was completed using lamp data acquired during the same night; we corrected each spectrum by the corresponding wavelength shift (which amounts, at
most, to 0.9 km s$^{-1}$). 

\section{Analysis methods and results: radial and rotational velocities} \label{sect_methods}

In \ion{Paper}{i}, we analyzed the stellar content of the NEP survey and we 
determined spectral types of all of the stellar counterparts to the 144
\textit{ROSAT} NEP X-ray sources. Our analysis confirmed the presence of an excess of active yellow stars, in
agreement with previous results for shallower surveys. 
To characterize the nature of the NEP stellar sample, we determine the radial velocity (RV) of each star, which provides a kinematic index of the population to which they belong, and the projected rotational velocity $vsini$, which is strongly correlated with the X-ray luminosity in single stars (Pallavicini et al. 1981), and evolves with stellar age (Skumanich 1972).

We derive simultaneously the radial and rotational velocities using the
Fourier Cross-correlation technique originally described by Tonry \& Davis (1979). The
cross-correlation technique was developed to derive robust radial velocities using all of the
information in the spectrum at once, rather than one spectral line at a time. 
We correlate the spectrum of a program star against that of a stellar
template. The position of the correlation peak indicates the relative radial velocity, and the
breadth of the peak is interpreted as being due to rotational broadening.

\begin{figure}
\centerline{\includegraphics[width=8cm]{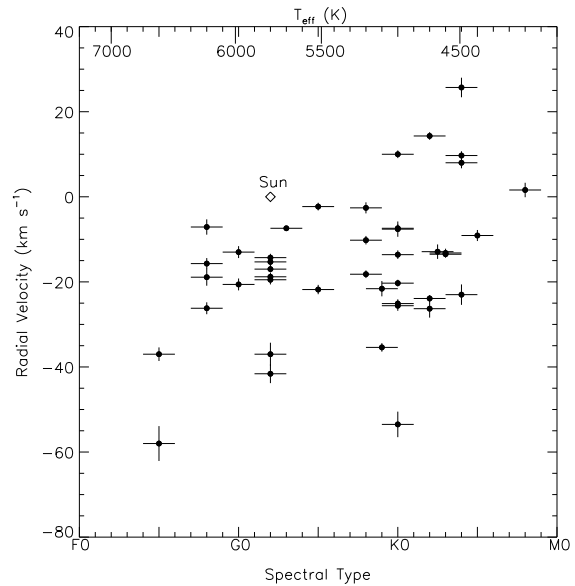}}
\caption{Heliocentric radial velocity distribution of the NEP
sample stars with respect to spectral type (temperature). We assumed that the
weighted mean of three independent measures is our estimate of radial velocity
for each star. The uncertainty was computed by combining in quadrature the
uncertainties of the measures.} 
\label{fig:9383fig1}
\end{figure}

\subsection{Radial velocities}
To check the robustness of our approach, we compared the cross-correlation method with that
performed line by line, then we calculated radial velocities adopting three different approaches which rely on:
\begin{enumerate}
\item{Wavelength shifts of selected photospheric absorption features}
\item{Cross-correlation function of the object
spectrum with that of a template star belonging to the sample}
\item{Cross-correlation function of the object
spectrum with that of the Sun, obtained with the same instrumental configuration as the target stars} 
\end{enumerate}

\begin{table}
\caption{Absorption features selected to calculate radial velocities from wavelength shifts}
\begin{center}
\label{tab1}
\begin{tabular}{cccc} \hline
Wavelength (\AA)	& Element & Wavelength (\AA)	& Element\\\hline
6419.95 & \ion{Fe}{i}& 6705.10 &\ion{Fe}{i}\\
6421.35 &\ion{Fe}{i}& 6710.32 &\ion{Fe}{i}\\
6430.85 &\ion{Fe}{i}& 6717.69 &\ion{Ca}{i}\\
6439.07 &\ion{Ca}{i}& 6726.67 &\ion{Fe}{i}\\
6449.81 &\ion{Ca}{i} &6743.12 &\ion{Ti}{i}\\
6450.23 &\ion{Co}{i} &6750.15 &\ion{Fe}{i}\\
6455.60 &\ion{Ca}{i} &6767.77 &\ion{Ni}{i}\\
6462.57 &\ion{Ca}{i} &6783.70 &\ion{Fe}{i}\\
6469.19 &\ion{Fe}{i}& 6786.86 &\ion{Fe}{i}\\
6471.66 &\ion{Ca}{i}& 6806.84 &\ion{Fe}{i}\\
6475.62 &\ion{Fe}{i}& 6810.26 &\ion{Fe}{i}\\
6481.87 &\ion{Fe}{i} &6814.95 &\ion{Co}{i}\\
6482.80& \ion{Ni}{i}& 6820.37 &\ion{Fe}{i}\\
6493.78 &\ion{Ca}{i} &6828.59 &\ion{Fe}{i}\\
6494.98 &\ion{Fe}{i}& 6839.83 &\ion{Fe}{i}\\
6495.74 &\ion{Fe}{i} &6841.34 &\ion{Fe}{i}\\
6496.47 &\ion{Fe}{i} &6842.04 &\ion{Ni}{i}\\
6496.90 &\ion{Ba}{ii}& 6842.69 &\ion{Fe}{i}\\
6569.22 &\ion{Fe}{i} &6843.66 &\ion{Fe}{i}\\
6572.78 &\ion{Ca}{i} &7090.38 &\ion{Fe}{i}\\
6574.23 &\ion{Fe}{i} &7095.43 &\ion{Fe}{i}\\
6575.02 &\ion{Fe}{i} &7110.90& \ion{Ni}{i}\\
6581.21 &\ion{Fe}{i} &7112.18 &\ion{Fe}{i}\\
6586.31 &\ion{Ni}{i}& 7122.20& \ion{Ni}{i}\\
6592.91& \ion{Fe}{i} &7130.92 &\ion{Fe}{i}\\
6593.87& \ion{Fe}{i} &7142.52& \ion{Fe}{i}\\
6598.60& \ion{Ni}{i} &7145.32 &\ion{Fe}{i}\\
6609.11& \ion{Fe}{i} &7148.15 &\ion{Ca}{i}\\
6703.57& \ion{Fe}{i} &7151.49& \ion{Fe}{i}\\\hline
\end{tabular}
\begin{flushleft}
\end{flushleft}
\end{center}
\end{table} 

\begin{figure}[h]
\centerline{\includegraphics[width=8cm]{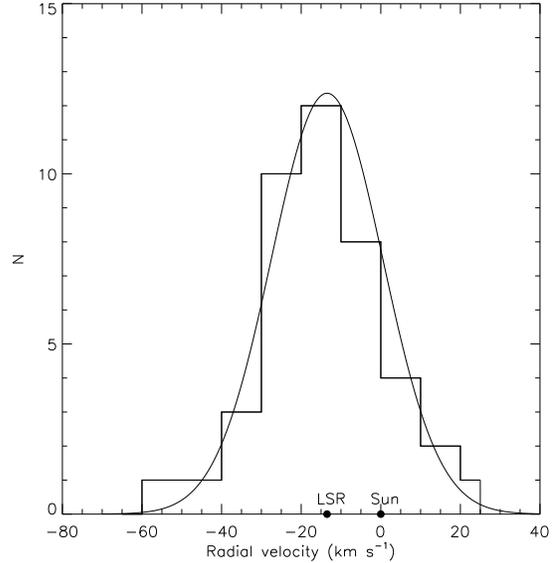}}
\caption{Histogram of the heliocentric radial velocities, the Sun velocity with
respect to the Local Standard of Rest (LSR) is indicated ($|v_{LSR}|$=13.4 km s$^{-1}$, Binney \& Merrifield, 1998). The superimposed gaussian radial velocity distribution for the targets with a standard deviation of 13 km s$^{-1}$, is shown.} 
\label{fig:9383fig2}
\end{figure}

We comment on the application of all three methods:
\begin{enumerate} 
\item{The heliocentric RVs of the sample stars were obtained by measuring the wavelength
shifts of a set of 58 measurable and unblended prominent photospheric absorption features (Table~\ref{tab1}), using the IRAF task {\it RVIDLINES}, and adopting the weighted mean of the results (using the inverse of the variances as weights). The variance of
the weighted mean is the inverse of the sum of the weights.}
\item{We calculated the RV of the sample stars using the IRAF task {\it FXCOR}, which performs a Fourier
Cross-Correlation of a given spectrum with a ``template" stellar spectrum, as
described by Tonry \& Davis (1979). For each star observed, we 
cross-correlated its spectrum with that of the template using the wavelength range between
about 6625 \AA\, and 6710 \AA, which contains no significant telluric absorption lines. While
performing the cross-correlations for individual stars, the wavelength ranges used were visually inspected and  obvious emission lines, cosmic rays, and/or bad pixels were masked. 
As our stellar template, we used a high signal-to-noise (S/N $\approx 180$) spectrum of the
bright star nep 6163 (V $\approx 7.6$), which has a modest projected rotational
velocity ($vsini \approx$ 3.7 km s$^{-1}$, derived by comparison with synthetic spectra
computed using ATLAS9). 
The heliocentric RV of the template object nep 6163 was obtained by measuring the wavelength
shifts of a list of prominent, photospheric, absorption features, as for method 1. The heliocentric RV
of each observed object is then calculated by adding the heliocentric RV of the template star, to the RV of each observed star, with respect to the template. The uncertainty in each value of heliocentric velocity was computed by combining (in quadrature) the uncertainties in the heliocentric RV of the
template star, and in the relative RV of the object, measured by cross-correlation
with the template star.}
\item{We performed an analysis similar to method 2, using as a stellar template, a high signal-to-noise (S/N $\approx$ 400) solar spectrum, acquired with the same instrumental configuration as the sample stars. Using IRAF task {\it RVCORRECT}, we find the heliocentric correction to the observed radial velocity of the solar spectrum, due to the relative motion Sun-Earth.
To measure the RV of each observed star, we cross-correlated each spectrum, with respect to the corrected template spectrum.}
\end{enumerate}

\begin{figure}[h]
\centerline{\includegraphics[width=8cm]{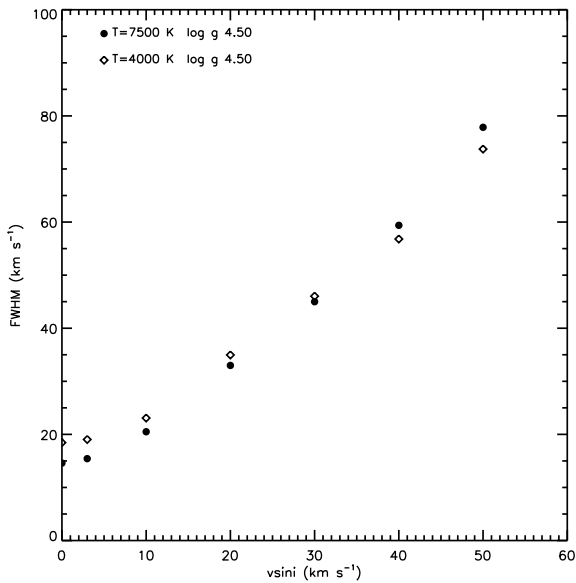}}
\caption{Calibration curve used for the $vsini$ estimation from the width of the
cross-correlation peak. The curve was created by cross-correlating the 
spectrum of the template star with a set of synthetic spectra with different rotational velocities.}
\label{fig:9383fig3}
\end{figure}

The three measurements agree to within a few km s$^{-1}$, the maximum difference being about 2.5 km s$^{-1}$; we therefore use the weighted mean, which is obtained using the inverse of the variances as weights, as our radial velocity measurement for each star; the variance that we associate with this weighted mean is the inverse of the sum of the weights.

For a few stars observed, the peak of the cross-correlation function
is asymmetric, probably due to binarity and/or high rotation rates ($\ge$ 60 km s$^{-1}$), particularly in early F-type stars;  we are unable however to determine whether the stars are
binaries and/or fast rotators because our spectra were obtained
during a single night. For about 7 stars, this asymmetry is indicative of a binary nature. These stars are discussed later. 
Figure \ref{fig:9383fig1} shows the heliocentric radial velocities as a function of spectral type for the whole sample.
The histogram in Fig. \ref{fig:9383fig2} provides the radial velocity distribution for single stars, which peaks significantly, with most velocities being in
the range between 0 and -30 km s$^{-1}$, and has a mean value of about -15 km s$^{-1}$. \\

\subsection{Rotational velocities}\label{3.2}
The cross-correlation function measures the projected rotational velocities ($vsini$) of our targets, relative to stellar spectra of essentially zero rotational velocity (our ``templates"); this is possible because the measured width of the cross-correlation peak is a function of the broadening of the spectrum lines, due to rotation velocity.
Assuming that the line broadening of the NEP stellar spectra is dominated by 
rotational broadening, and that our stellar template is an intrinsically narrow-lined spectrum, the width of the cross-correlation peak is a measure of the $vsini$ of the NEP targets, which may be calibrated as discussed below. 
We use a standard Local Thermodynamic Equilibrium (LTE) spectral analysis with the latest version of the
spectral-line analysis code MOOG (Sneden 1973), and a grid of ATLAS9 model atmospheres (Kurucz 1993), to compute synthetic spectra for comparison with the observed spectra.

The ATLAS9 stellar atmospheres are computed
without the overshooting option, and by assuming a mixing length to pressure scale height
ratio of $\alpha$=0.5. Assumptions made in the models include: the atmosphere is
plane-parallel and in hydrostatic equilibrium, the total flux is constant, the
source function is described by the Planck function, and the populations of different
excitation levels and ionization stages are governed by LTE. The atomic data
used to generate synthetic spectra are taken from {\it Kurucz Atomic Line
Database} (Kurucz 1995).  

The cross-correlation is performed with {\it FXCOR}. The main peak of the cross-correlation
function is fit using a Gaussian function and its FWHM is measured. These FWHM values are converted to $vsini$
values, using the following calibration process.
We compute a grid of synthetic spectra, assuming $\log
g$=4.5 (cm s${^{-2}}$),  microturbulent velocity $\xi$=1.0 km s$^{-1}$ and solar
metallicity, which are reasonable parameters for our template star, assuming the
SARG instrumental resolution, R=57000. We convolved these spectra with a
rotational broadening function with $vsini$ varying from 0 to 50 km s$^{-1}$. We allowed
$T_{\rm eff}$ to vary between 4000 K and 7500 K (which are the typical
temperature values for our targets) in 500 K intervals, to
check the sensitivity of this relationship to temperature.

The synthetic spectra were cross-correlated with our template and the FWHM of the
cross-correlation peak was measured. We compared the results derived for
the same analysis using synthetic spectra computed for $\log
g$=3.0 (cm s${^{-2}}$). The comparison highlights that the calibration results
are almost independent of gravity, while the dependence on temperature
is more pronounced.

The resulting relation between the FWHM of the cross-correlation peak and the corresponding $vsini$, for the lowest ($T_{\rm eff}$ = 4000 K) and the highest ($T_{\rm eff}$ = 7500 K) temperature values, are shown in Fig. \ref{fig:9383fig3}. 

\begin{figure}
\centerline{\includegraphics[width=8cm]{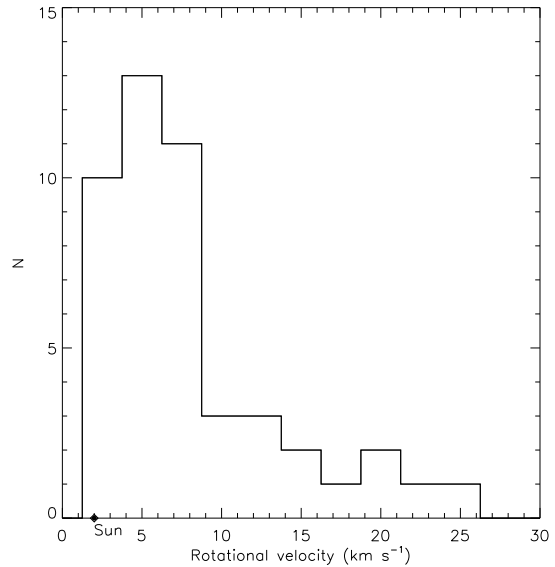}}
\caption{Histogram of the rotational velocities; almost all of the sample stars have
rotational velocities that are comparable or higher than that of the Sun ($\approx$2 km s$^{-1}$).}
\label{fig:9383fig4}
\end{figure}

We note that the discrepancies between the calibration curves, calculated for different $T_{\rm eff}$, are consistent within the typical errors inferred for the
rotational velocities measured for our targets. Using
the relationship corresponding to the expected temperature of our template (according to spectral type), we linearly
interpolated the FWHM of the cross-correlation peak, with the derived calibration
curve to estimate the $vsini$ values. \\The errors of the rotational velocities were
calculated as prescribed by Rhode, Herbst \& Mathieu (2001), to be $v
sini/(1+r)$, where $r$ is the Tonry \& Davis (1979) parameter, which is a
measure of the signal-to-noise ratio of the cross-correlation peak.\\ 
When the calibration derived is applied to determine $vsini$ for each star, we
must pay attention to the minimum value of $vsini$ to which this procedure is sensitive.
The measured width of the cross-correlation function can be described by the
following formula (Tonry \& Davis, 1979):

$$
\sigma_{meas}^2=\sigma_{rot}^2+\sigma_{nat}^2+2\sigma_{inst}^2
$$
where $\sigma_{rot}$ is the rotational width, $\sigma_{nat}$ is the natural, or intrinsic, line
width which is negligible, and $\sigma_{inst}$ is the instrumental broadening (the last term of the equation includes two contributions because two spectra form the cross-correlation function).

For the derivation of the
minimum detectable $vsini$, we follow Bailer-Jones (2004). The terms are additive; 
with perfect data of ``infinite" S/N ratio, we could therefore determine ``any" rotational broadening, even with
non-zero instrumental broadening. In practice, this is not however possible due to noise. Bailer-Jones (2004) assumes that $\sigma_{rot}$ is detectable only if its value exceeds that of other broadening contributions, and its minimum measurable value is $\sqrt{2}\sigma_{inst}$.
Our theoretical FWHM of the instrumental broadening is fixed by the slit width to be 5 km s$^{-1}$. We note
that the {\it full} width of a rotational profile corresponds to twice the
rotational velocity: one half of the line is created by the blueshifted
approaching limb of the star, the other half by the redshifted receding limb
(Bailer-Jones, 2004). The minimum detectable $vsini$ for our data is 
$\sqrt{2}$ x 5.0/2 = 3.5 km s$^{-1}$, which corresponds to the flat portion of the calibrations in Fig. \ref{fig:9383fig3}. 

The cross-correlation technique measures the projected, rotational
velocity of the Sun to be 1.63$\pm$0.34 km s$^{-1}$. However, this measurement, which agrees with the well-known value assumed for the Sun
($\approx$\, 1.86 km s$^{-1}$, Soderblom 1982), is below the minimum detectable $vsini$ of our study. As we can see from Fig. \ref{fig:9383fig3}, this value appears in the flat portion of the
calibration relation; the adoption of the mean instrumental broadening as the upper limit on the
$vsini$ for the slowest rotators, is therefore conservative. \\We indicate the $vsini$ values derived for slow rotators with $\le $ 3.5 km s$^{-1}$.\\

To perform a further comparison, we adopted a second method to calculate rotational
velocities.\\ 
We performed a detailed comparison between the observed and synthetic
spectra in the wavelength range 6420 \AA\, - 6480 \AA, which includes
several Fe I and Ca I absorption lines, of moderate strength ($\approx
$ 100 m\AA). Using Kurucz model atmospheres and the MOOG software, we generated a
grid of LTE synthetic spectra. We allowed the stellar temperature
vary, and assume $\log g$=4.5 (cm s${^{-2}}$), $\xi$=1.0 km s$^{-1}$, a solar
metallicity, and a macroturbulent velocity, 
that is constrained by the spectral type (Fekel, 1997); the rotational velocity was then the only free
parameter. Each spectrum was finally convolved by a 0.1 \AA/px Gaussian FWHM,
which reproduces the 
instrumental profile, as estimated by our lamp spectra. For the determination of rotational velocities, we used the neutral iron lines at 6430.85 \AA, 6546.24 \AA\, and 6569.22 \AA\, and the \ion{Ca}{i} lines at 
6439.07 \AA\, and 6471.66 \AA, which were strong but not saturated lines, and unblended with other lines. 
We compared the observed and synthetic spectra and
determined by eye the value of $vsini$ that provides the most appropriate fit to the observed target rotational velocity. 
We assumed that the spread in $vsini$ values for individual lines represented the uncertainty in each measurement. We note that the $vsini$ values were measured by comparing observed spectra and synthetic profiles that had been convolved by all broadening effects. The minimum value of $vsini$ to which this procedure is sensitive is therefore formally zero.
The two independent measurements agree to within a few km s$^{-1}$; the maximum difference is
about 3.1 km s$^{-1}$ (for the low S/N spectra). We therefore 
calculated the weighted mean of the two measurements of $vsini$ and the inverse of the sum of the weights was taken to be the variance associated with the weighted mean. Figure \ref{fig:9383fig4} shows the histogram of the
rotational velocities.\\

\section{Binary stars}
\begin{figure}[h]
\centerline{\includegraphics[width=9cm]{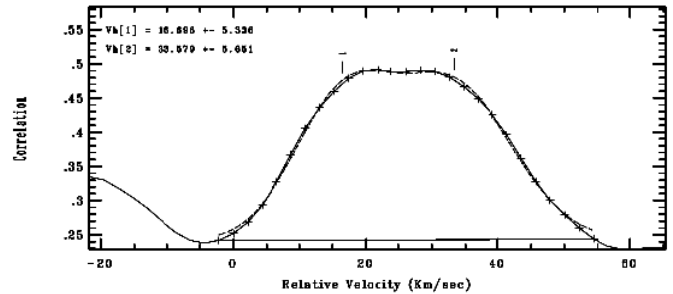}}
\caption{Example of peaks of the cross-correlation function produced by cross-correlating a binary with our template star.
Relative radial velocities of the two components are indicated in the panel.}
\label{fig:9383fig5}
\end{figure}

For approximately 7 stars, $\approx$\,12\% of all stars observed, the shape of the {\it FXCOR} cross-correlation function peak was asymmetric; this may be indicative of a binary nature for these stars. The peak of the
cross-correlation function can be fitted using two overlapping Gaussians, instead of one single function, when the blending between the two stars is not severe; an example of such a case can be seen in Fig. \ref{fig:9383fig5}. However, we cannot measure the radial velocity of binary
systems, because our spectra were acquired during a single night. Both radial and rotational velocities for such stars, when measurable, refer to the contributions of both stars. We report these results for completeness
and stress that they are affected by blending of both contributions, and must therefore be taken with
extreme caution. The only exception is the spectroscopic
binary (SB2) nep 5520, for which spectral lines from both components are clearly separated and the
calculations refer separately to each star. For this system, the two components appear to have similar spectral
types and effective temperatures, i.e. to contribute equally to the continuum.\\

\begin{figure}
\centerline{\includegraphics[width=8cm]{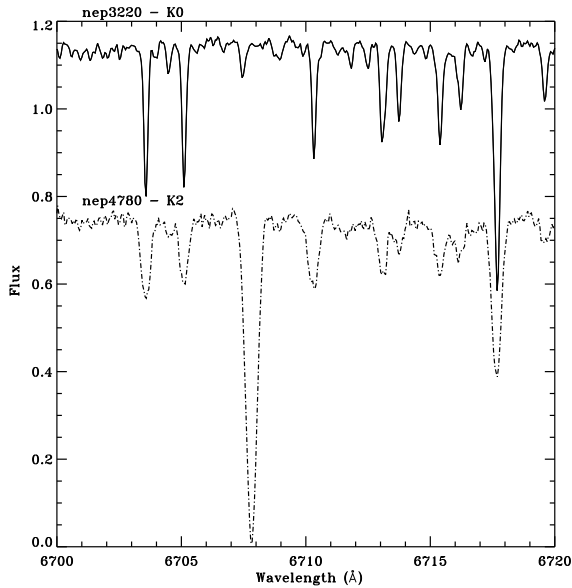}}
\caption{Spectra in the lithium region for two NEP sources: nep 3220 does not show the lithium line while nep 4780
shows a
prominent line. The spectra have been continuum subtracted and offset on the
vertical axis.}
\label{fig:9383fig6}
\end{figure}

\section{Lithium abundances}
Of the 56 total observed stars, 25 show a prominent lithium feature, 5 of which are supposed to be binaries.
For these stars, we used the IRAF task {\it Continuum} to perform the spectrum normalization. We created
a synthetic model atmosphere, using the ATLAS9 code (Kurucz 1993), assuming $\log g$=4.5 (cm
s${^{-2}}$), $\xi$=1.0 km s$^{-1}$ and solar
metallicity, and derived the effective temperatures using the star's spectral type. 

These models were used
to create synthetic spectra for a small interval of approximately 50 \AA\, about the lithium feature at
6707.8 \AA, from 6700 \AA\, to 6750 \AA, to 
determine the line-free regions that were fitted using low-order cubic spline functions, and a variable residual
rejection limit. Both the line-free spectral range and the rejection limit were chosen interactively in each case, based on the visual inspection of fitting results. In Fig. \ref{fig:9383fig6},
we show the normalized spectra about the \ion{Li}{i} 6707.8 \AA\, line for two stars observed: the spectrum of  nep 3220 does not show the lithium line while nep 4780 shows a prominent line. The \ion{Li}{i} 6707.8 \AA\, line is close to the \ion{Fe}{i} 6707.44 \AA\, line and both lines are often blended, due to spectral line broadening by
stellar rotation. The equivalent width of each of the partially-blended lines, \ion{Li}{i} 6707.8 \AA\,
and \ion{Fe}{i} 6707.44 \AA, was determined using a two-Gaussian fitting routine provided by IRAF. To estimate uncertainties in the EW measurements, we repeated both the continuum normalization and EW measurement
three times. We then accepted the weighted mean of the three measures to be final EW measurements and
as typical error the standard deviation of errors, which were typically not more than a few percent ($\approx$ 3 m\AA).

For stars with 4000 K ${\le}$$T_{\rm
eff}$${\le}$6500 K, the Li abundances N(Li) were derived using the measured EWs and from a linear interpolation of the growth curves of Soderblom et al. (1993). Effective temperatures were derived from the spectral type (Allen, 1973), as determined in \ion{Paper}{i}; we derived the error in two subtypes from the published spectral types (\ion{Paper}{i}), and this uncertainty was propagated into the Li abundances.

\begin{figure}[h]
\centerline{\includegraphics[width=8cm]{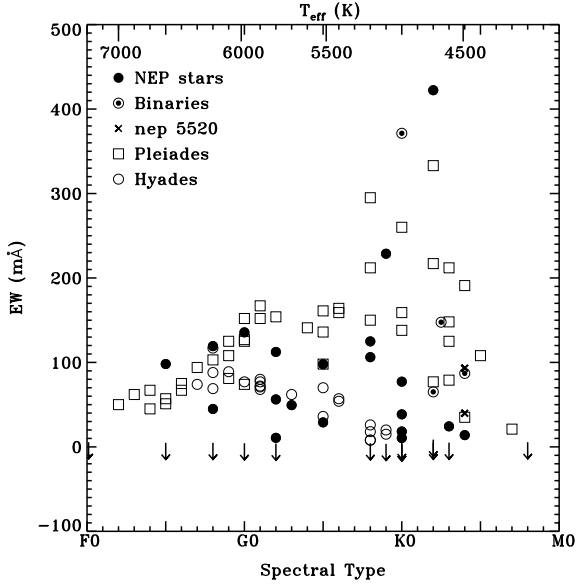}}
\caption{Distribution of lithium equivalent widths of single NEP stars (filled
dots) and binary stars (ringed
dots; '$\times$' for nep 5520), with respect to spectral
type, compared to Pleiades (empty squares) and Hyades (empty dots) distributions.
Arrows indicate EW upper
limits for those stars in which no lithium line was visible in the spectrum.}
\label{fig:9383fig7}
\end{figure}

The lithium EW for binary stars was computed in relation to the combined continuum, which was the
sum of continuum contributions from both components. Without information about the individual spectral types
of both components, we were unable to separate the continuum contributions and obtain
individual estimates of the lithium EW of each star (apart from the SB2 nep 5520).
We measured the EWs for the suspected binary stars, but we 
emphasize that these measurements correspond to the unsolved blend including the contribution of both stars; we can therefore only state that these stars have lithium. While this is the best one
can do with the available information, this procedure is likely to introduce additional errors. Therefore in
Fig. \ref{fig:9383fig7}, in which we report the lithium EWs measured for the subsample of single
stars, such system are flagged with a different symbol (filled squares, while the EWs for the two
components of nep 5520 are indicated as '$\times$'). 

In Table~\ref{tabB}, we
indicate the Li abundances N(Li) derived, together with the NLTE corrections interpolated from the grid of Carlsson et al. (1994). 
When no lithium line was visible in the
spectrum, we used the statistical noise measured from the spectral continuum, to determine an upper limit to the Li line EW, except for those stars with such high rotational velocity, that the lines were completely smeared out. 
To set constraints on the age of our targets, the lithium EWs of the Pleiades
(10${^8}$ yrs) and Hyades (10${^9}$ yrs) clusters are plotted in
Fig. \ref{fig:9383fig7} for comparison (Soderblom et al., 1993 and 1990,
respectively).

The comparison of the three samples of stars shows that the EWs of most targets
lie between those of stars of similar temperature in the two clusters, and in a few cases are lower than those of the Hyades.

\begin{table}
\caption{Properties of high radial velocity stars}
\begin{center}
\label{tab2}
\begin{tabular}{rcccc} \hline
nep  &  Sp.Type & High Vrad 	&High Vrot &Li\\
&&($\ge |30|$ km s$^{-1}$)&($\ge 15$ km s$^{-1}$)&\\\hline
1690 &  G2&-37.0$\pm$2.7 & - &-\\
1800 &  F5&-58.0$\pm$4.1 & 16.3$\pm$1.4 &3.30$\pm$0.20\\
3560 &  G8&-26.0$\pm$1.3 & - &2.30$\pm$0.30\\
*4090 & G2& -41.6$\pm$2.2&- &-\\
4810 & F5& -37.0$\pm$1.6 & - &-\\
5510 & K0& -53.5$\pm$3.0 & 20.4$\pm$2.1 &3.70$\pm$0.50\\\hline
\end{tabular}
\begin{flushleft}
 *: See NOTE in Table~\ref{tabA}
\end{flushleft}
\end{center}
\end{table}

\section{Discussion}
We derive both radial and rotational velocities, and lithium abundances, for the stellar population of
the NEP survey. 

We use different methods to determine the same parameters,
which enables us to infer the reliability of the analysis techniques employed. We obtain results that
are consistent with each other, which demonstrated the robustness of our analysis. We quote the results in Table~\ref{tabB}: col.1 identifies stars with NEP numbers; col.s 2-3
provide the heliocentric RV and the $vsini$ values, with estimated errors (we indicate that 3.5 km s$^{-1}$ is the upper limit for measurements of rotational velocity that are below the instrumental profile width, as discussed in Sect. \ref{3.2}); cols.4-7 indicate the equivalent
widths of the lithium line (6707.8 \AA), the temperature and its associated error (derived from spectral type), and lithium abundances (derived from the interpolation of the growth curves of Soderblom
et al., 1993), respectively; col.8 reports the
spectral types and in col.9 we indicate whether the star is a binary (label ``x") or a double-line spectroscopic
binary (label ``SB2"). Label ``*" indicates that we lack the Thorium-Argon lamp exposures for the nights in which data for these stars were acquired. These observations
were wavelength-calibrated using lamp exposures acquired for the night closest in time. We assumed the same spectral type and temperature for the two component of SB2
nep 5520; this is an ``ad hoc" assumption, which is
reasonable because the lines are similar. In Table~\ref{tabB}, we indicate our poor knowledge about binaries with the label ``?", close to the spectral type indication.

\begin{figure}
\centering
\includegraphics[width=8cm]{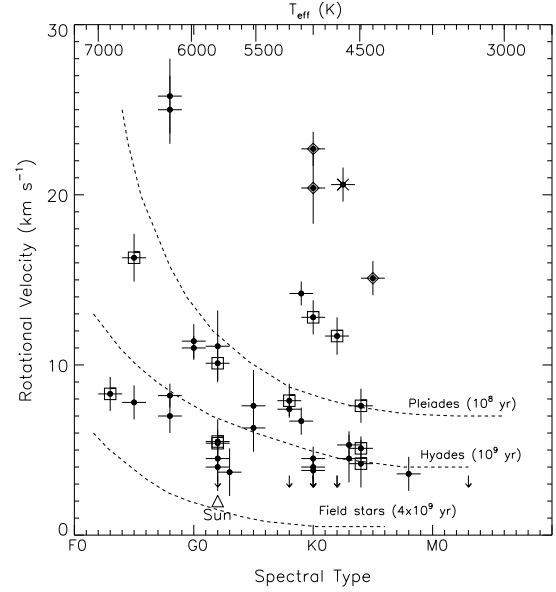}
\caption{Rotational velocity distribution with respect to spectral type
(temperature); average trends of rotational velocity of Pleiades, Hyades and
field stars, are shown (Favata et al. 1995; Kraft 1967; Soderblom 1983a,b; Skumanich 1972). With squares we indicate stars that have {\ion{Na}{i}}($D_1$,$D_2$) emission; ``$\times$" indicates
the presence of H$\alpha$\, emission and diamonds are for stars
that have both H$\alpha$\, and {\ion{Na}{i}}($D_1$,$D_2$) emission. }
\label{fig:9383fig8}
\end{figure}

\subsection{Radial velocity}
In Figure \ref{fig:9383fig2}, we show the histogram of the heliocentric radial velocities, and 
superimpose a Gaussian radial velocity distribution for the target stars, adopting a standard
deviation of 13 km s$^{-1}$, that corresponds to the histogram variance, and a mean value of -13.5 km s$^{-1}$, that agrees well with the bulk of the data. 

The Sun moves
relative to the Local Standard of Rest (LSR) with a mean velocity of
$|v_{LSR}|$=13.4 km s$^{-1}$ (Binney \& Merrifield, 1998).

The Gaussian radial velocity distribution, which represents a good fit for most of 
the targets reflects the motion of the Sun relative to the LSR. The distribution of radial velocity is consistent with that of a young population
of disk stars of age 4x10${^9}$ yrs (Wilson 1963; Kraft 1967; Skumanich 1972; Simon 1985), or younger.
We detect a small number (6) of stars with large radial velocity ($\ge
|30|$ km s$^{-1}$). Nevertheless we cannot consider these stars as a statistical meaningful sample, because only 4 of these stars are within
2$\sigma$ from the mean value, one is within 3$\sigma$ and only one has a velocity which falls beyond 3$\sigma$ from the mean value with an extremely small probability
($|P|>$\, 3 $\sigma$ $\approx$ 0.3\%).

In Table~\ref{tab2}, we report the properties of these stars and the lithium abundances. In
particular we want to know if these stars are older than the
remainder of the observed stars or if they are peculiar in some
way. Of these six stars, three of them (nep 1690, nep 4090, nep 4810) are slow rotators
($\le 15$ km s$^{-1}$) that have no detected 6708 \AA\,
lithium line. These could be old stars. One of these stars (nep 3560) is a slow
rotator and a Li line is detectable in its spectrum, and finally,
two stars (nep 1800, nep 5510) are fast rotators ($\ge 15$ km s$^{-1}$) and have measurable Li line in their spectra.

A possible explanation for the ``high velocity" stars can be given in terms of binarity. Nep 5510 is probably a binary system. Analyses of spectra acquired for all other stars do not, however, confirm conclusively the presence of binarity.

Abundance analyses, in particular the measurement of the metallicity and chemical abundance ratios of several elements ([$\alpha$/Fe], [O/Fe]), could provide insight into the stellar populations, because abundances are a key indicator of which population a star belongs to.

\subsection{Rotational velocity}
Figure \ref{fig:9383fig4}
shows that almost all stars have projected rotational velocities, which are
interpreted as lower limits to actual values, that are comparable to or higher than that of the
Sun ($\approx 2$ km s$^{-1}$). 

Rotational velocity decreases with increasing age. Since magnetic
activity depends on rotation and differential rotation, the level of magnetic activity and
hence X-ray emission will be different for fast and slow rotators. 

Our sample seems to be dominated by
relatively young or intermediate-age stars. This statement is supported by Fig. \ref{fig:9383fig8},
which shows the $vsini$ distribution as a function of the spectral type and compares it to the average trends for the
Pleiades, Hyades and field stars; we indicate stars that have {\ion{Na}{i}}($D_1$,$D_2$) emission with square symbols, H$\alpha$\, emission with $\times$ symbols, and stars
that have both H$\alpha$\, and {\ion{Na}{i}}($D_1$,$D_2$) emission with diamond symbols. 
We see that most stars have measurements that
are consistent with an age between 10${^8}$ yrs (Pleiades age) and
4x10${^9}$ yrs (field stars), compared to the age of the Sun, 5x10${^9}$ yrs.

A small number of K stars have measured
rotational velocities that are higher than measurements for the Pleiades. We propose that these stars have ages that are comparable to those of the Pleiades, or are even younger. 
Two of the stars could be binaries. For the remainder of the sample the youthness is the most plausible explanation.

Widely-separated binaries behave in a similar way to single stars because tidal forces can be neglected (Slettebak 1963, Weiss 1974). Close binaries, however, show altered rotation. Close binaries, of early spectral
type, rotate more slowly than single stars, while those of later type rotate much faster than
their single counterparts (Gray 1976, 1984).

\begin{figure}[h]
\centerline{\includegraphics[width=8cm]{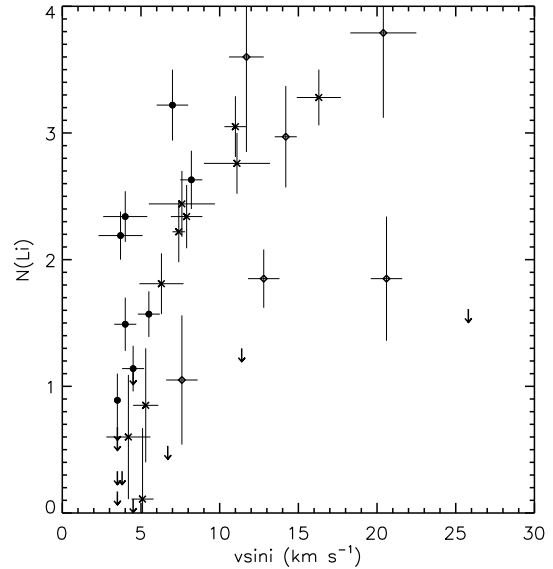}}
\caption{Lithium abundances as a function of the rotational velocity: diamonds indicate
stars with $vsini$ greater than those of Pleiades; ``x" indicate $vsini$ values
between the average curves of Pleiades and Hyades; dots indicate $vsini$ values
between the average curves of Hyades and field stars; down-pointing arrows indicate upper limits of lithium
abundances.}
\label{fig:9383fig9}
\end{figure}

\begin{figure}[h]
\centerline{\includegraphics[width=8cm]{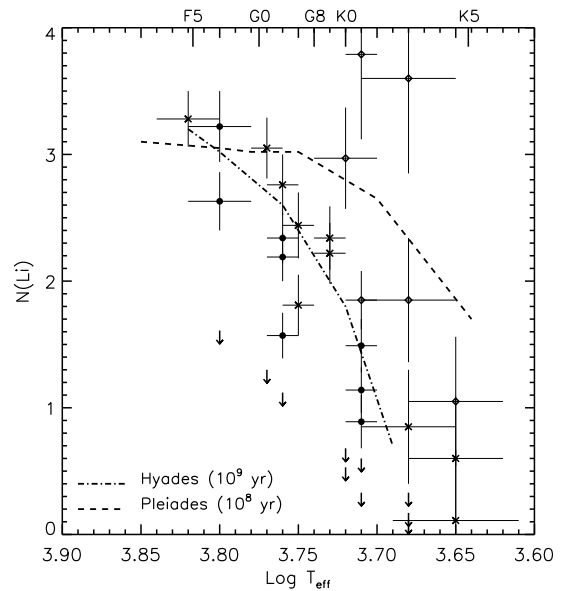}}
\caption{Lithium abundances as a function of the effective temperatures: symbols as
in Fig. \ref{fig:9383fig9}. We indicate Pleiades and Hyades fits, obtained 
by Favata et al. (1993).}
\label{fig:9383fig10}
\end{figure}

In
Figures \ref{fig:9383fig9} and \ref{fig:9383fig10}, we report the LTE lithium abundances, upper limits included, as a function of rotational
velocity and temperature, respectively.
Stars that show high lithium abundances and low rotation are probably young; this hypothesis is
reinforced by the presence of emission in the 
H$\alpha$ and {\ion{Na}{i}} $D{_1}$ and $D{_2}$ lines, in most spectra (see Table~\ref{tabC}).

\subsection{Comparison with model results} 
The predictions of the model XCOUNT (Favata et al. 1992; Micela et al. 1993) for the complete NEP survey, for each
spectral type and age range are summarized in Table~\ref{tab3}.

XCOUNT assumes an exponential
disk-like spatial distribution of the stars, with a radial scale length of 3.5 kpc, as in Bahcall \& Soneira
(1980). We followed the modifications of the Bahcall \& Soneira (1980) model of Micela et al. (1993), which
introduced an age-dependent scale height, dividing stars into three age ranges: 10$^7$-10$^8$, 10$^8$-10$^9$,
and 10$^9$-10$^{10}$ years, with scale heights of 100, 200 and 400 pc, respectively. Micela et al. (1993) assumed X-ray luminosity functions derived from \textit{ROSAT} observations of the Pleiades (Micela et al. 1996) and Hyades (Stern et al. 1995), and from \textit{Einstein} data for nearby stars (Schmitt et al. 1985; Maggio et al. 1987; Barbera et al. 1993), which are considered to be prototypes for the three age ranges.

We scaled the XCOUNT predictions by a factor corresponding to the fraction of the stars that we observed ($\approx$ 0.6 for late-F to G stars and $\approx$ 0.4 for K stars), and assigned each star to a given age range based on its lithium EWs and/or rotational velocities. We used lithium as a primary indicator and assigned a star: to the  10$^7$-10$^8$ years age range, if its lithium EWs were greater than (or
comparable to) those of the Pleiades; to the 10$^8$-10$^9$ years age range, if its lithium EWs were between the measurements shown by the average curves of Pleiades and Hyades; and to the 10$^9$-10$^{10}$ years age range, if its lithium EWs were less than those of the Hyades. For those stars for which lithium is not detected, we used rotational velocities as an alternative indicator, applying the same criteria as for the lithium EWs. We focus on late-F to G and K type stars because our age indicators (lithium and rotation) are not appropriate for early-type stars. 
For late-F to G stars, we expect to find about 1.2 stars that have an age $\le$\,10$^8$ yrs, but instead find 7 stars. Similarly, we expect about 5 stars to have an intermediate value of age in the range
10$^8$-10$^9$ years, and indeed we find 4 stars. 
For the age range 10$^9$-10$^{10}$ years, we find 6 stars in contrast to about 5 predicted by the model. 
For K stars, we find 7 stars that have an age $\le$\,10$^8$ yrs, compared with about 2 stars predicted by the model; 5 stars to have an intermediate age, between
10$^8$-10$^9$ years, with respect to about 4 predicted stars; and finally 7 stars in the range
between 10$^9$-10$^{10}$ years, in contrast to a prediction of 8 stars. Our results are summarized in Table~\ref{tab4}: we indicate the observed and predicted (by XCOUNT model) stars for our sample, and,
between brackets, the total counts of stars that may have an age in the range considered, on the basis of rotational velocities and lithium EWs, including those stars with only rotational velocity
and no lithium measurements. For 8 of the late-F to G and K stars observed, we did not measure the upper limits of lithium EWs due to the low S/N ratio of the spectra and/or high rotation.

The observed excess seems to be concentrated at young age. Our comparison 
suggests a burst of star formation, with an increasing of new-born stars by a factor of $\approx$ 4.5 (or $\approx$ 3.5 if we exclude 4 probable binary systems, concentrated at young age) in the last 10$^8$ years, while it is consistent with a constant stellar birthrate in the intermediate and old ages, between 10$^8$-10$^{10}$ years.

\begin{table}
\caption{Summary of X-ray source counts predictions for each spectral type and
range of age derived from XCOUNT (from \ion{Paper}{i}).}
\begin{center}
\label{tab3}
\begin{tabular}{lcccc} \hline\hline
Sp.Type	&	N(young)&N(intermed.)&	N(old)&Total\\\hline
A&	0.63&1.41	&0.28&2.32\\
F0-F5&	0.32&1.53	&7.34&9.19\\
F6-F9+G&1.91&8.42	&7.63&17.96\\
K&	4.24&8.86	&20.26&33.36\\\hline
&7.1	&20.22	&35.51&62.83\\\hline
\end{tabular}
\begin{flushleft}
\end{flushleft}
\end{center}
\end{table} 

\begin{table*}
\caption{Summary of F6-F9, G and K observations derived in this work vs
scaled predictions derived from XCOUNT, for each 
range of age. For 8 stars of our sample of late-F to G and K stars we did not measure lithium EWs (upper limits) due to the low S/N ratio of their spectra and/or high rotation.}
\begin{center}
\label{tab4}
\begin{tabular}{lcccccccc} \hline\hline
\multicolumn{1}{l}{Sp.Type}&\multicolumn{2}{c}{N(young)}&\multicolumn{2}{c}{N(intermed.)}&\multicolumn{2}{c}{N(old)}&\multicolumn{2}{c}{Total}\\\hline
       &pred. & obs. &  pred.  & obs.   & pred. & obs.& pred. & obs.\\\hline
F6-F9+G&1.15  & 7 (9)&  5.05   & 4 (5)  & 4.58  & 6 (8) &   10.78 & 17 (22)\\
K      &1.70  & 7 (10)&  3.54   & 5 (5)    & 8.10  & 7 (7)     &    13.34 & 19 (22)\\\hline
       &2.85  & 14 (19)&8.59   & 9 (10) &12.68  & 13 (15)&   24.12 & 36 (44)\\\hline
\end{tabular}
\begin{flushleft}
\end{flushleft}
\end{center}
\end{table*}

\section{Conclusions}
We derived fundamental quantities, such as radial and
rotational velocities and lithium abundances, for stars detected in the NEP X-ray survey.
In \ion{Paper}{i}, we confirmed the presence of an excess of yellow stars in the NEP sample and
suggested that such an excess could be due to the presence of a young
population, formed by a recent event of star formation. Moreover, there was no evidence for an equivalent
excess of M stars, which would be necessary to identify
the population as really young. We refer the reader to \ion{Paper}{i} for a possible explanation of the absence of an excess. In the present work, M stars are poorly sampled ($\approx $ 0.05\% of the entire sample of NEP M stars) and, as a consequence, we cannot state anything about their mean properties.
Our results, which refer to F, G and K stars, confirm that most of the stars 
seem to belong to a young age population. The population excess of the model prediction described in \ion{Paper}{i} appears to be composed of mainly young stars (with ages comparable to those of Pleiades). 

Our data suggest a burst in the birthrate, by a factor of $\approx$ 4, in the last 10$^8$ years. The Gould Belt is an asymmetry in the apparent distribution of the brightest stars in the sky, with respect to Galactic equator, probably associated with a recent star formation episode close to the Sun (Guillout et al 1998). The NEP surveyed area 
is far from the Gould Belt and this star formation episode was not accounted for in the XCOUNT model, and is possible that other local, less prominent events of star formation occurred in the solar neighborhood. 
Nevertheless few stars, among the fastest K rotators, may be old binary systems with tidally locked rotation.\\

\begin{acknowledgements}
Based on observations made with the Italian Telescopio Nazionale Galileo (TNG) operated on the island of La Palma by
the Fundaci\`{o}n Galileo Galilei of the INAF (Istituto Nazionale di Astrofisica) at the Spanish Observatorio del Roque
de los Muchachos of the Instituto de Astrofisica de Canarias. 
We acknowledge support by the Marie Curie Fellowship
Contract No. MTKD-CT-2004-002769 and ASI-INAF I/088/06/0 Contract. This research has made use of the SIMBAD database, operated at CDS, Strasbourg, France,
and the NASA's Astrophysics Data System Abstract Service. We wish to thank the referee M. Guedel for his careful reading of the manuscript and for
his suggestions to improve the paper. We wish to thank Javier Lopez Santiago for fruitful
discussions.
\end{acknowledgements}

\appendix
\section{Observing log for the NEP spectra and results}

In this section, we present the observing log for the observations of stars detected in NEP survey (hereafter NEP spectra) and the results derived during the present study. \\In Table~\ref{tabA}, we report the observing
log for the NEP spectra, including identification numbers,
observation dates, right ascension, declination (Eq. 2000.0), spectral types (as
derived from Micela et al. 2007), and total observing times.\\ In
Table~\ref{tabB}, we provide our measurements: heliocentric RV and $vsini$
values, with estimated errors, equivalent
widths of lithium line (6707.8 \AA), temperatures with associated errors (derived from spectral type), and
lithium abundances (derived from the interpolation of the growth curves of Soderblom
et al., 1993).\\In Table~\ref{tabC}, we report the presence of H$\alpha$\, and
{\ion{Na}{i}}($D_1$,$D_2$) emission, and of a measurable lithium line at 6707.8 \AA.
 
\begin{table*}
\caption{Observing Log for the NEP spectra. Scan numbers are from Gioia et al. 2003.}
\begin{center}
\label{tabA}
\begin{tabular}{rcccccc} \hline
    
Scan    &  Date   &    Ra	&     Dec  &      Spectral type&     $t_i$ (s)  & S/N \\ 
&&(h m s)&$^\circ~'$  ${}^{\prime\prime}$&&\\\hline

  *20 &  2004 Jun 28 & 17 52 44.8 &+67 00 20&       G8  &  5400&70\\
 260 &  2004 Jun 01 & 17 58 28.3 &+67 26 08&	  G2&	2400&50\\	
 330 &  2004 Jun 02 & 17 59 23.6 &+66 02 55&	  G9&  1800&70\\
 370 &  2004 May 25 & 18 00 02.2 &+66 45 53&	  F8&  2400&90\\
*1420 &  2004 Mar 08 & 17 19 28.8 &+65 22 29&	  K0&  3600&60\\
*1441 &  2004 Jan 31 & 17 20 05.6 &+62 06 22&	  G2&  1200&100\\
1470 &  2004 Mar 09 & 17 20 26.7 &+67 03 37&	  G8&  1800&80\\
1500 &  2004 Apr 07 & 17 21 10.9 &+69 48 02&	  G2&  600&200\\
1511 &  2004 Feb 03 & 17 21 42.5 &+62 00 32&	  K4&  3600&60\\
1580 &  2004 Apr 07 & 17 24 00.4 &+69 40 30&	  K2&  3600&50\\
*1600 &  2004 Mar 08 & 17 24 26.8 &+64 12 23&	  G2&  1800&50\\
1601 &  2004 Mar 09 & 17 24 38.8 &+64 40 51&	  G0&  1800&80\\
1690 &  2004 Mar 09 & 17 26 45.4 &+69 37 53&	  G2&  2400&60\\
1800 &  2004 May 06 & 17 29 39.4 &+68 47 38&	  F5&  1200&110\\
1840 &  2004 May 25 & 17 30 19.9 &+69 55 27&	  G0&  1800&90\\
1960 &  2004 Apr 07 & 17 33 16.9 &+67 12 08&	  F2&  600&170\\
2050 &  2004 Mar 09 & 17 36 14.1 &+65 02 27&	  K0&  2400&70\\
2100 &  2004 May 06 & 17 36 26.6 &+68 20 37&	  M3&  600&170\\
2210 &  2004 Jun 02 & 17 39 16.1 &+70 20 09&	  F8&  1200&80\\
2250 &  2004 Mar 08 & 17 39 56.1 &+65 00 04&	  K0&  2400&200\\
2360 &  2004 Jun 01 & 17 42 26.5 &+69 07 58&	  F8&  1200&120\\
2400 &  2004 Mar 09 & 17 43 01.6 &+66 06 46&	  G2&  1800&90\\
2470 &  2004 Jun 02 & 17 43 51.7 &+70 31 39&	  F3&  1800&40\\
2480 &  2004 Jun 01 & 17 44 00.6 &+70 15 27&	  K0&	2400&50\\
*2970 &  2004 Jan 31 & 17 49 03.9 &+62 47 48&	  F2&  600&290\\
*3080 &  2004 Jun 28 & 17 50 25.3 &+70 45 36&	  K2.5&  1800&90\\
3220 &  2004 Jun 01 & 17 52 56.0 &+66 25 10&	  K0&	600&150\\
3560 &  2004 Jun 28 & 17 57 03.7 &+68 49 14&	  G8&  1800&110\\
3610 &  2004 May 06 & 17 58 01.4 &+64 09 34&	  K4&  2700&70\\
3690 &  2004 May 06 & 17 58 48.0 &+63 50 39&        A2&  1200&120\\
3710 &  2004 Jun 30 & 17 59 13.8 &+64 08 33&	  F2&	1200&200\\
*4090 &  2004 Jun 28 & 18 05 30.5 &+69 45 17&	  G2&  2400&50\\
4100 &  2004 May 25 & 18 05 30.3 &+62 19 03&	  K0&  1200&150\\
4130 &  2004 Jun 02 & 18 05 44.8 &+65 51 58&	  F5&  1200&80\\
4180 &  2004 May 25 & 18 06 41.0 &+64 13 18&	  K0&  600&440\\
4380 &  2004 Jun 01 & 18 08 45.4 &+62 56 37&        G5&	1200&120\\
4470 &  2004 Jun 29 & 18 09 55.8 &+69 40 39&	  K3&  1200&160\\
4530 &  2004 Jun 29 & 18 10 49.9 &+70 16 09&	  G9&  4800&80\\
4780 &  2004 Jun 29 & 18 13 48.6 &+68 31 32&	  K2&  2400&80\\
4810 &  2004 Jun 03 & 18 13 51.4 &+64 23 57&	  F5&  180&300\\
*4931 &  2004 Jun 28 & 18 16 21.2 &+65 29 39&	  K8&  1200&170\\
4970 &  2004 Jun 03 & 18 16 49.7 &+65 04 26&	  K2&  2400&60\\
5060 &  2004 Jun 29 & 18 18 31.9 &+70 42 17&	  G5&  1800&90\\
5220 &  2004 Jun 03 & 18 20 19.2 &+65 19 19&	  F8&  600&120\\
*5320 &  2004 Jun 28 & 18 21 46.8 &+63 57 10&	  K0&  1800&90\\
5510 &  2004 Jun 03 & 18 25 09.0 &+64 50 21&	  K0&  600&260\\
5520 &  2004 Jun 03 & 18 25 32.9 &+62 34 15&	  K4&  2400&60\\
5950 &  2004 Jun 30 & 18 32 29.5 &+68 36 52&	  G0&  600&190\\
6030 &  2004 Jun 29 & 18 33 47.8 &+65 13 33&	  K3&  1200&150\\
6051 &  2004 Jun 29 & 18 34 33.6 &+69 31 45&	  K5&  3600&50\\
6160 &  2004 Jun 30 & 18 36 13.5 &+65 29 15&	  F0&  300&230\\
6163 &  2004 Jun 29 & 18 36 22.8 &+66 54 54&	  G3&  600&180\\
6350 &  2004 Jun 29 & 18 39 25.4 &+69 02 54&	  M3&  2400&60\\
6400 &  2004 Jun 29 & 18 40 44.4 &+70 38 47&	  K2&  1200&70\\
6410 &  2004 Jun 30 & 18 40 56.6 &+62 44 54&	  K0&  300&300\\
6510 &  2004 Jun 29 & 18 43 12.6 &+69 55 54&	  F0&  1200&80\\\hline
\end{tabular}
\begin{flushleft}
*: We lack the Thorium-Argon lamp exposures for the nights in which these stars were acquired. They
were wavelength calibrated using lamp exposures of the nearest night.\\
\end{flushleft}
\end{center}
\end{table*}

\begin{table*}[h]
\caption{RESULTS: Vrad, Vrot, Li equivalent width, Li abundance, Binarity}
\begin{center}
\label{tabB}
\begin{tabular}{rrrcccccc} \hline
    
nep    &  Vrad (km s$^{-1}$)   &    Vrot (km s$^{-1}$)	&     $EW_{\rm Li}$(m\AA)  &   $T_{\rm eff}(K)$ &$N_{\rm Li}$ & $N_{\rm Li}$$^{\rm NLTE}$ &Spectral type &notes \\ \hline
  *20       &  -17.3$\pm$0.9  & $\le$3.5      &$<$\,4.3   & 5310$\pm$150	 & $<$\,0.68        &$<$\,0.77	      &  G8  &\\
 260        &  -19.5$\pm$1.1  & 11.1$\pm$2.1  &112.4     & 5790$\pm$150		&2.76$\pm$0.24	   &2.71$\pm$0.19   & G2&\\	
 330        &  -35.4$\pm$1.0  & 6.7$\pm$0.8   &$<$\,3.6   & 5250$\pm$150	 &$<$\,0.53	    	&		& G9&\\
 370        &  -15.7$\pm$1.3  & 7.0$\pm$1.0   &119.3     & 6250$\pm$300		&3.22$\pm$0.28     &3.09$\pm$0.20	& F8&\\
*1420       &                 & 22.7$\pm$1.0  &-         & 5150$\pm$150		&  		  	&		&K0&\\
*1441       &  -15.7$\pm$1.5  & 4.0$\pm$1.4   &56.2      & 5790$\pm$150		&2.34$\pm$0.20	   &2.34$\pm$0.18	& G2&\\
1470        &  -10.2$\pm$1.0  & 7.4$\pm$0.4   &106.2     & 5310$\pm$150		& 2.22$\pm$0.24	   & 2.27$\pm$0.21	& G8&\\
1500        &  -17.0$\pm$0.8  & 5.5$\pm$0.7   &10.7      & 5790$\pm$150		&1.57$\pm$0.18     &1.59$\pm$0.18	& G2&\\
1511        &                 &              &87.0      & 4450$\pm$350		&1.01$\pm$0.50    	&		& K4 (?)&x\\
1580        &                 &              &65.1      & 4830$\pm$300		& 1.36$\pm$0.40   & 1.55$\pm$0.37	& K2 (?)&x\\
*1600       &  -18.8$\pm$0.9  & 5.4$\pm$1.4   &-         & 5790$\pm$150		& 		   	&		& G2&\\
1601        &  -13.0$\pm$1.4  & 11.0$\pm$0.7  &135.6     & 5940$\pm$150		& 3.05$\pm$0.24    & 2.94$\pm$0.18	&G0&\\
1690        &  -37.0$\pm$2.7  & 10.1$\pm$1.0  &-         & 5790$\pm$150		&  		   	&		&G2&\\
1800        &  -58.0$\pm$4.1  & 16.3$\pm$1.4  &98.1      & 6650$\pm$300		& 3.28$\pm$0.22	   & 3.16$\pm$0.18	 &F5&\\
1840        &                 &    $\ge$30.0           &-         & 5940$\pm$150		&  		  	&		 &G0&\\
1960        &                 & $\ge$50.0     &-         & 7000$\pm$300		 &  		   	&		&F2&\\
2050        &  -7.6$\pm$1.8   & 12.8$\pm$1.0  &77.2      & 5150$\pm$150		 & 1.85$\pm$0.23   & 1.96$\pm$0.21	& K0&\\
2100        &                 &              &-         & 3500$\pm$350		&  		   	&		&M3&\\
2210        &  -7.1$\pm$1.8   & 25.0$\pm$2.0  &-         & 6250$\pm$300		 & 		    	&		& F8 (?)&x\\
2250        &  -7.4$\pm$0.9   & 3.8$\pm$0.7   &$<$\,3.0   & 5150$\pm$150	  &$<$\,0.33 	      	&		& K0&\\
2360        &  -26.2$\pm$1.4  & 8.2$\pm$0.7   &44.9      & 6250$\pm$300		 & 2.63$\pm$0.23   & 2.58$\pm$0.16	&F8&\\
2400        &  -14.3$\pm$0.9  & 4.5$\pm$1.4   &$<$\,3.9   & 5790$\pm$150	  &	$<$\,1.12    &	$<$\,1.16 	& G2&\\
2470        &                 & 8.3$\pm$1.0   &-        & 6950$\pm$300		 &  		    	&		& F3&\\
2480        &                 & $\ge$40.0             &-         & 5150$\pm$150		 & 		    	&		& K0&\\
*2970       &                 & $\ge$60.0     &-         & 7000$\pm$300		 & 		    	&		& F2 (?)&x\\
*3080       &  -12.9$\pm$1.7  & 20.6$\pm$1.0  &147.6     & 4800$\pm$350		 & 1.85$\pm$0.49   & 2.00$\pm$0.53	& K2.5 (?)&x\\
3220        &  -25.1$\pm$1.0  & $\le$3.5      &$<$\,5.4   & 5150$\pm$150	  & $<$\,0.60         	&$<$\,0.72		&K0&\\
3560        &  -2.6$\pm$1.3   & 7.9$\pm$1.0   &124.9     & 5310$\pm$150		 &  2.34$\pm$0.25  &  2.38$\pm$0.24	& G8&\\
3610        &  +9.7$\pm$1.0   & 5.1$\pm$0.7   &13.9      & 4450$\pm$400		 &   0.11$\pm$0.56   	&		&K4&\\
3690        &                 &              &-         &  9000$\pm$800 	&		    	&		&A2&\\
3710        &                 &$\ge$30.0      &-         & 7000$\pm$300 	 &		     	&		&F2&\\
*4090       &  -40.7$\pm$2.2  & $\le$3.5      &-         & 5790$\pm$150 	 & 		     	&		&G2&\\
4100        &  +10.0$\pm$0.9  & 4.5$\pm$0.7   &18.2      & 5150$\pm$150		 & 1.14$\pm$0.18    & 1.27$\pm$0.31	 &K0&\\
4130        &                 &$\ge$40.0      &-         & 6650$\pm$300		 &  		     	&		&F5&\\
4180        &  -13.6$\pm$1.0  & $\le$3.5      &$<$\,3.0   & 5150$\pm$150	  & $<$\,0.33 		&		&K0&\\
4380        &  -21.8$\pm$1.1  & 6.3$\pm$1.4   &29.0     &   5560$\pm$200         & 1.81$\pm$0.24    & 1.86$\pm$0.23	&G5&\\
4470        &  -13.5$\pm$0.8  & 5.3$\pm$0.8   &24.3     & 4800$\pm$350		 & 0.85$\pm$0.45    & 1.06$\pm$0.44	 &K3&\\
4530        &  -21.6$\pm$1.8  & 14.2$\pm$0.7  &228.7    & 5300$\pm$150		 & 2.97$\pm$0.40    & 2.84$\pm$0.26	&G9&\\
4780        &  -26.3$\pm$2.1  & 11.7$\pm$1.1  &422.3    & 4830$\pm$300		 &  3.60$\pm$0.75   &  3.40$\pm$0.65	& K2&\\
4810        &  -37.0$\pm$1.6  & 7.8$\pm$1.0   &-        & 6650$\pm$300		 & 		      	&		& F5&\\
*4931       &  +2.5$\pm$1.7   & 3.6$\pm$1.0   &$<$\,5.5  & 4000$\pm$150		  & $<$\,-0.8		&		& K8&\\
4970        &  +14.3$\pm$0.9  & $\le$3.5      &$<$\,8.6  & 4830$\pm$300		  &$<$\,0.41 		&		& K2&\\
5060        &  -2.3$\pm$0.9   & 7.6$\pm$2.1   &97.9     & 5560$\pm$200		 & 2.44$\pm$0.26    & 2.45$\pm$0.22	&G5&\\
5220        &  -18.9$\pm$2.0  & 25.8$\pm$2.2  &$<$\,4.8  & 6250$\pm$300 	  &$<$\,1.61 		&$<$\,1.59 	&F8&\\
*5320       &  -19.4$\pm$0.8  & 4.0$\pm$0.7   &38.5     & 5150$\pm$150		 & 1.49$\pm$0.21     & 1.62$\pm$0.20	&K0&\\
5510        &  -53.5$\pm$3.0  & 20.4$\pm$2.1  &371.3    & 5150$\pm$150		 & 3.79$\pm$0.67     & 3.46$\pm$0.54	&K0 (?)&x\\
5520${^a}$  &   +8.0$\pm$1.3    & 7.6$\pm$1.0   &93.2     & 4450$\pm$350		  & 1.05$\pm$0.51	&		& K4 (?)&SB2\\
5520${^b}$  &   +25.7$\pm$2.3   & 4.2$\pm$1.4   &39.8     & 4450$\pm$350		  & 0.60$\pm$0.49	&		& K4 (?)&\\
5950        &  -20.6$\pm$1.4  & 11.4$\pm$1.0  &$<$\,4.3  & 5940$\pm$150		   & $<$\,1.30 		& $<$\,1.31	&G0&\\
6030        &  -13.2$\pm$1.0  & 4.5$\pm$1.4   &$<$\,5.0  & 4780$\pm$350		   &$<$\,0.11  		&		&K3&\\
6051        &  -9.1$\pm$1.3   & 15.1$\pm$1.0  &-        & 4410$\pm$400		  &  			&		&K5&\\
6160        &                                &  &-     & 7300$\pm$300		 & 			&		& F0&\\
6163        &  -7.4$\pm$0.5   & 3.7$\pm$1.4   &49.4     & 5700$\pm$150		  &2.19$\pm$0.19	&2.21$\pm$0.18	&G3&\\
6350        &                 & $\le$3.5      &-        & 3500$\pm$350		  &  			&		&M3&\\
6400        &  -23.9$\pm$0.8  & $\le$3.5      &$<$\,5.0  & 4830$\pm$300		   & $<$\,0.17		  &		& K2&\\
6410        &  -25.6$\pm$1.2  & $\le$3.5      &10.5     & 5150$\pm$150		  &0.89$\pm$0.21 	&1.02$\pm$0.21 	& K0&\\
6510        &                 &              &-        & 7300$\pm$300		 & 		        &		& F0&\\\hline
\end{tabular}
\begin{flushleft}
NOTE: An x in column eight indicates that the star is (or it is supposed to be) binary. We indicate SB2 for double-lined spectroscopic
binary.\\ *: We lack the Thorium-Argon lamp exposures for the nights in which these stars were acquired. They
were wavelength calibrated using lamp exposures of the nearest night.\\
${^a}$: primary and ${^b}$: secondary component of nep 5520 SB2 system.\\
\end{flushleft}
\end{center}
\end{table*}

\begin{table*}
\caption{Presence of H$\alpha$\, and
{\ion{Na}{i}}($D_1$,$D_2$) emission and presence of lithium feature (6707.8 \AA)}
\begin{center}
\label{tabC}
\begin{tabular}{rcccc} \hline
    
nep  &binary     &    H$\alpha$	&     Na I ($D_1$,$D_2$)  &      Li I\\ \hline

 *20 &   & -  &- &      -  \\
 260 &   & -  &- &      x\\	
 330 &   & -  &- &      -  \\
 370 &   & -  &- &      x  \\
*1420 &   & x  &x &     -  \\
*1441 &   & -  &- &     x  \\
1470 &   &-  &- &       x  \\
1500 &   & -  &x &      x  \\
1511 &x   &-  &x &      x  \\
1580 &x   & x  &x &     x  \\
*1600 &x   & -  &x &    -  \\
1601 &   & -  &- &      x  \\
1690 &   & -  &x &      -  \\
1800 &   & -  &x &      x  \\
1840 &   & -  &- &      -  \\
1960 &   & -  &- &      -  \\
2050 &   & -  &x &      x  \\
2100 &   & -  &- &      -  \\
2210 &x   & -  &- &     -  \\
2250 &  & -  &- &       -  \\
2360 &   & -  &- &      x  \\
2400 &   & -  &- &      -  \\
2470 &   & -  &x &      -  \\
2480 &   & x  &x &      -  \\
*2970 &   & -  &- &     -  \\
*3080 &x   & x  &- &    x  \\
3220 &   & -  &- &      -  \\
3560 &   & -  &x &      x  \\
3610 &   & -  &x &      x  \\
3690 &   & -  &- &      -  \\
3710 &   & -  &x &      -  \\
*4090 &  & x  &x &      -  \\
4100 &   &-  &- &       x  \\
4130 &   & -  &- &      -  \\
4180 &   & -  &- &      -  \\
4380 &   &-  &- &       x   \\
4470 &   &-  &- &       x   \\
4530 &   & -  &- &      x  \\
4780 &   & -  &x &      x  \\
4810 &   & -  &- &      -  \\
*4931 &   &-  &- &      -  \\
4970 &   &-  &x &       -  \\
5060 &   & -  &- &      x  \\
5220 &  & -  &- &       -  \\
*5320 &   & -  &- &     x  \\
5510 &x  &x  &x &       x   \\
5520 &SB2   & -  &x &   x\\
5950 &   & -  &- &      -  \\
6030 &   & -  &- &      -  \\
6051 &   & x  &x &      -  \\
6160 &   & -  &- &      -  \\
6163 &   &-  &- &       x   \\
6350  &  & x  &x &      -  \\
6400 &   & -  &- &      -  \\
6410 &   & -  &- &      x \\
6510 &  & -  &- &       - \\\hline
\end{tabular}
\begin{flushleft}
*: We lack the Thorium-Argon lamp exposures for the nights in which these stars were acquired. They
were wavelength calibrated using lamp exposures of the nearest night.\\
\end{flushleft}
\end{center}
\end{table*}

\end{document}